\def\NAME#1#2{\caption{#2} \label{#1}}
\def\FIG#1{\begin{figure}[!htb] \begin{center} #1 \end{center} \end{figure}}
\def\FIGT#1{\begin{figure}[!t] \begin{center} #1 \end{center} \end{figure}}

\def\EPSF#1#2#3#4{\includegraphics[#3= #4 cm,clip]{#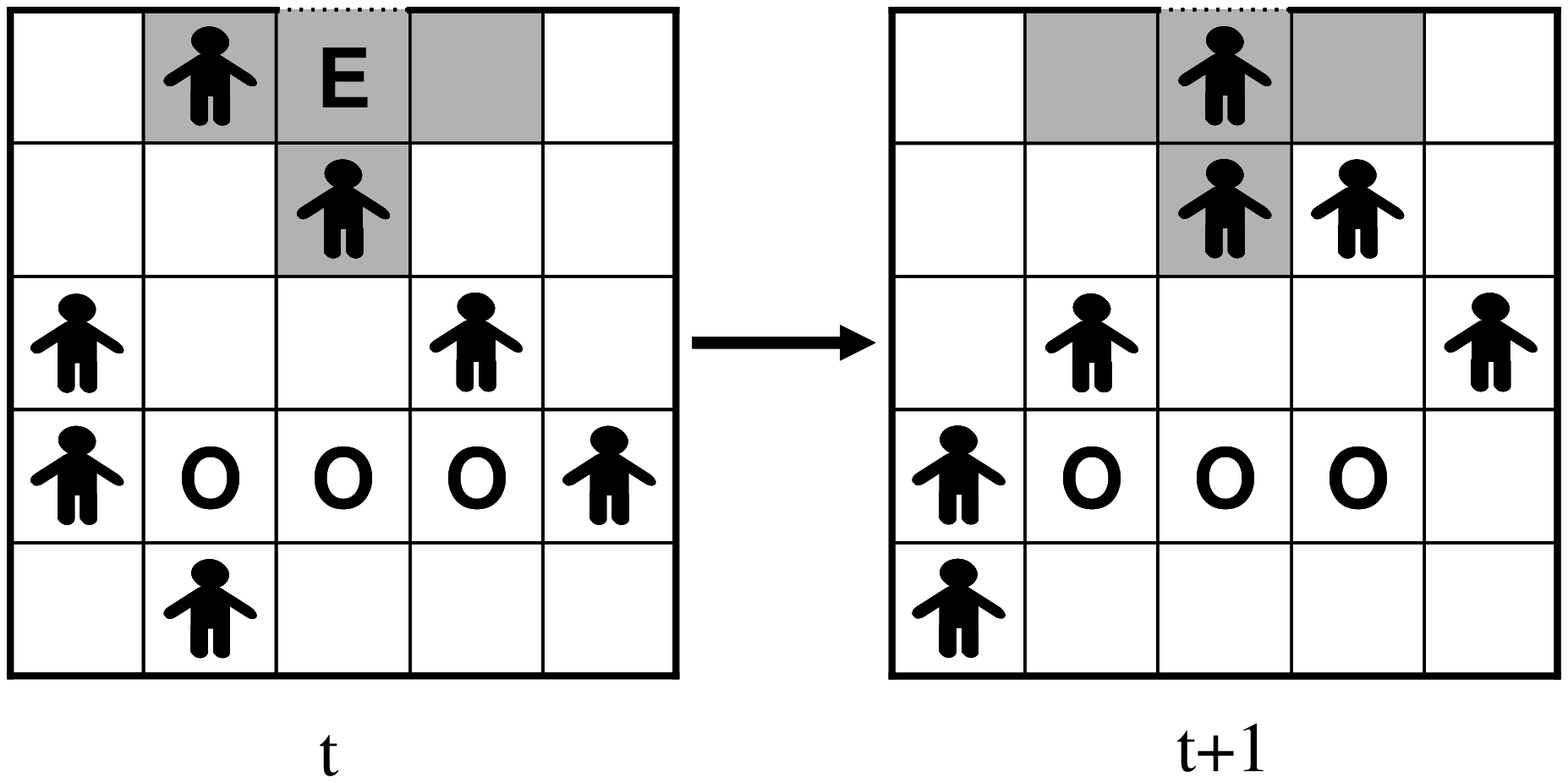} \NAME{#1}{#2}}

\def\ING#1#2#3{\includegraphics[#2=#3cm,clip]{#1.eps}}

\def\SEC#1#2{\section{#1} \label{#2}}

\documentclass[unsortedaddress,twocolumn,pre,amsmath,amssymb]{revtex4}
\usepackage[dvips]{graphicx} 
\usepackage{dcolumn} 
\usepackage{bm}

\usepackage{fancyhdr}
\usepackage{amsmath}

\begin{document}

\title{Mean Field Theory for Pedestrian Outflow through an Exit}
\author{Daichi Yanagisawa and Katsuhiro Nishinari }
\affiliation{%
Department of Aeronautics and Astronautics,
School of Engineering, University of Tokyo,
Hongo, Bunkyo-ku, Tokyo 113-8656, Japan.
}

\begin{abstract}
An average pedestrian flow through an exit is one of the most important index in evaluating pedestrian dynamics. 
In order to study the flow in detail, the floor field model, which is a crowd model by using cellular automaton, is extended by taking into account a realistic behavior of pedestrians around the exit. 
The model is studied by both numerical simulations and cluster analysis to obtain a theoretical expression of an average pedestrian flow through the exit. 
It is found quantitatively that the effect of exit door width, a wall, and pedestrian's mood of competition or cooperation significantly influence the average flow.
The results show that there is suitable width of the exit and position according to pedestrian's mood.
\end{abstract}

\maketitle

\SEC{Introduction}{Intro}

Pedestrian dynamics has received growing interest over the last decades from physicists since it shows new collective behaviors such as dynamical phase transitions and spontaneous symmetry breaking \cite{social, socialnature, socialreview, p-matrix}.
Helbing has designed the social force model \cite{socialnature} which reproduces typical pedestrian behavior such as an arching, lane formation, and oscillations of the direction at bottlenecks.
It is based on a system of coupled differential equations which have to be solved by using e.g. molecular dynamics approach similar to the study of granular matter.
Another approach is a discrete modeling using cellular automaton, which has been actively studied in recent years \cite{p-matrix, multigrid, largeobject, different-v}.
In this paper, we study the floor field (FF) model which is a cellular automaton model, introducing two kinds of FFs, i.e., Static FF (SFF) and Dynamic FF (DFF), to move pedestrians from one cell to another.
The two FFs, which are explained in the Sec.\ref{FFM}, enable us to simulate egress processes from complex rooms of arbitrary geometry quite efficiently.
Kirchner \textit{et al} discovered that pedestrian's mood of competition or cooperation increase or decrease the evacuation time \cite{competition}. Moreover an obstacle in front of the exit will shorten the evacuation time in some cases in the simulation of egress processes \cite{friction}.
Many extended models are proposed up to now to make the FF model more realistic.
For example, the strength of inertia of pedestrians which suppresses quick changes of the direction of motion is considered in \cite{inertia-wall}.
Henein \textit{et al} has taken into account physical forces between pedestrians by adding a dynamic force field to the FF model \cite{force}.

Most of these studies are based on simulations and there are few analytical results because of the complexity of rules of motion and two-dimensionality.
In this paper we present an analytical result on outflow through an exit, which is one of the most important index in evaluating evacuation dynamics.
Kirchner \textit{et al} obtained an expression of the average number of evacuated persons $\langle N \rangle$ from an exit with 1 cell's width as a function of time step and the \textit{friction parameter} $\mu$ by mean-field approximation \cite{friction}.
Here we newly introduce the \textit{bottleneck parameter} $\beta$ which makes pedestrian behaviors around the exit more realistic. 
We have succeeded to calculate the average flow $\langle Q \rangle$ as a function of $\beta$, $\mu$, and the width of an exit door $w$ by cluster approximation.
As far as we know, the analytical expression of the average flow through the exit with arbitrary width is derived for the first time in this paper.

This paper is organized as follows. 
In Sec.\ref{FFM} we briefly review the FF model and introduce the new parameter $\beta$ in Sec.\ref{BETA}.
We calculate the average flow $\langle Q \rangle$ by cluster approximation in Sec.\ref{CLUSTER}, and $\langle Q \rangle$ obtained from simulation and the theoretical expression are compared in Sec.\ref{S&E}.
In Sec.\ref{C&C} and Sec.\ref{WALL} we consider how the mood of the pedestrians and a wall beside the exit influence the average flow.
The both effects are explained by the contour plots of the average flow in Sec.\ref{CONTOUR}. 
Sec.\ref{CONC} is devoted to summary and discussion. 

\SEC{Floor Field Model}{FFM}

\subsection{Floor Field}

We consider a situation that every pedestrian in a room moves to the same exit.
The room is divided into cells as given in Fig.\ref{1}.
Man shaped silhouettes represent pedestrians, an alphabet \textbf{E} and alphabets \textbf{O} represent the exit cell and obstacle cells, respectively.
Each cell contains only a single pedestrian at most.
Every time step pedestrians choose which cell to move from 5 cells: a cell which the pedestrian stands now ($(i,j)=(0,0)$) and the Neumann neighboring cells ($(i,j)=(0,1)$, $(0,-1)$, $(1,0)$, $(-1,0)$) (Fig.\ref{2}).
Two kinds of FFs determine the probability of which direction to move.
SFF $S_{ij}$, which is the shortest distance to the exit cell, is given by the $L^2$ norm as
\begin{equation} \label{sffdf}
S_{ij}=\sqrt{|x_{ij}-x_{exit}|^2+|y_{ij}-y_{exit}|^2},
\end{equation}
where $(x_{ij},y_{ij})$ and $(x_{exit},y_{exit})$ are the coordinates of the cell $(i,j)$ and the exit cell respectively.
However, when there is an obstacle on the way to the exit, SFF is calculated by making detour of it (Fig.\ref{3}).
Thus, SFF is not simply described by (\ref{sffdf}) in the case \cite{inertia-wall}.
Pedestrians move to a cell that has smaller SFF than a cell they occupy and hence go to the exit.
DFF $D_{ij}$ at cell $(i,j)$ is a number of footprints left by the pedestrians.
Pedestrians interact with each other by footprints like ants do by their pheromone.
The long-ranged interaction by pedestrian's sight is approximated to short-ranged interaction around the pedestrian by using DFF.
This shortens calculation time dramatically.
When pedestrians move to the common exit, it is known in reality that they tend to follow each other.
This phenomenon can be reproduced by moving pedestrians to a cell which has bigger DFF (Fig.\ref{4}).
DFF has its own dynamics, namely diffusion and decay, which leads to broadening, dilution and finally vanishing of the footprints \cite{inertia-wall}.

Therefore in this model, the transition probability $p_{ij}$ for a move to a neighbor cell $(i,j)$ is determined by the following expression,
\begin{equation}
p_{ij}=N\xi _{ij}\exp (-k_sS_{ij} +k_{d}D_{ij}). \label{pp}
\end{equation}
Here the values of the FFs $S_{ij}$ and $D_{ij}$ at each cell $(i,j)$ are weighted by two sensitivity parameters $k_s$ and $k_d$ with the normalization $N$.
There is a minus sign before $k_s$ since pedestrian move to a cell which SFF decreases.
$\xi_{ij}$ returns 0 for an obstacle or a wall cell and returns 1 for other kinds of cells.
Note that in our paper a cell occupied by a pedestrian is not regarded as an obstacle cell, thus it affects the normalization $N$.

\FIG{\EPSF{1}{A schematic view of an evacuation simulation by the FF model. Pedestrians proceed to the exit by one cell at most by one time step.}{height}{3.3}}
\FIG{\EPSF{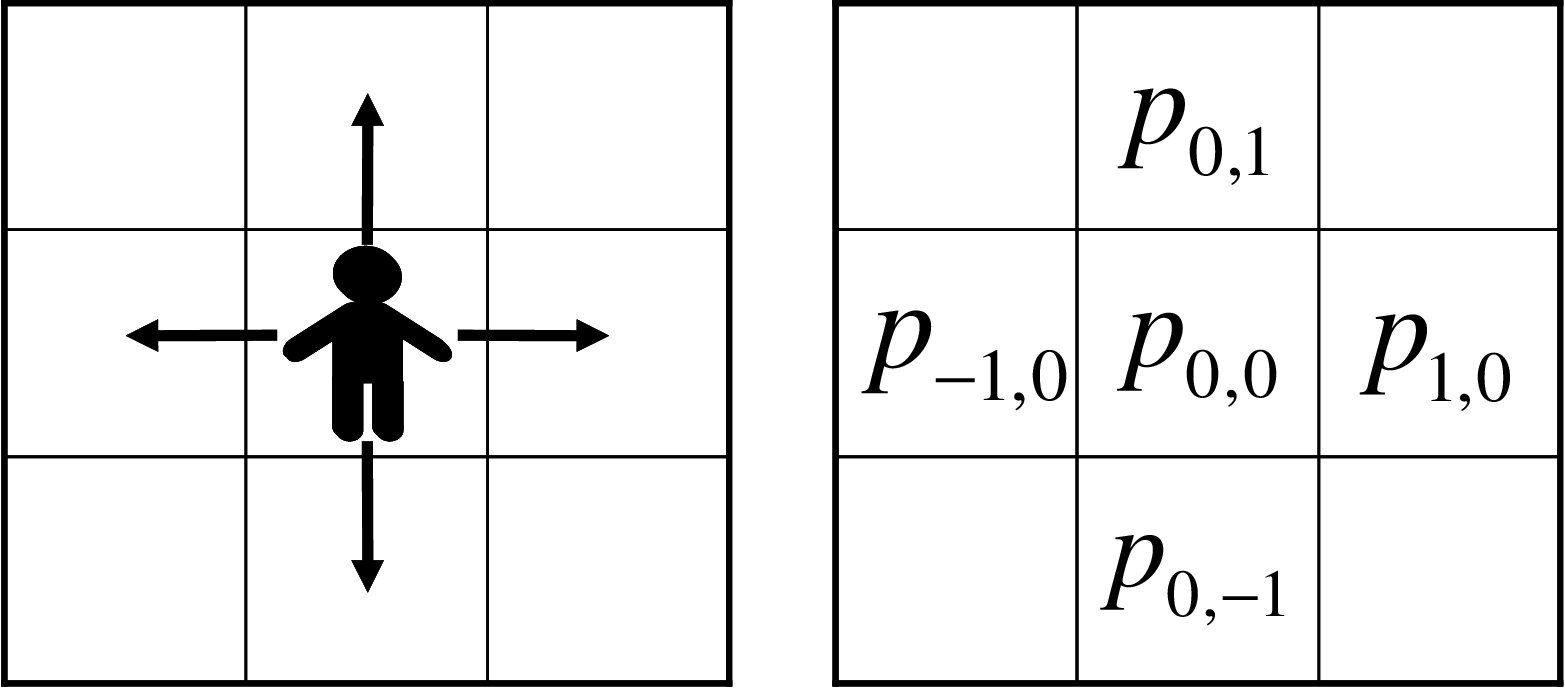}{Target cells for a pedestrian at the next time step. The motion is restricted to the Neumann neighborhood in this model.}{height}{2.5}}
\FIG{\EPSF{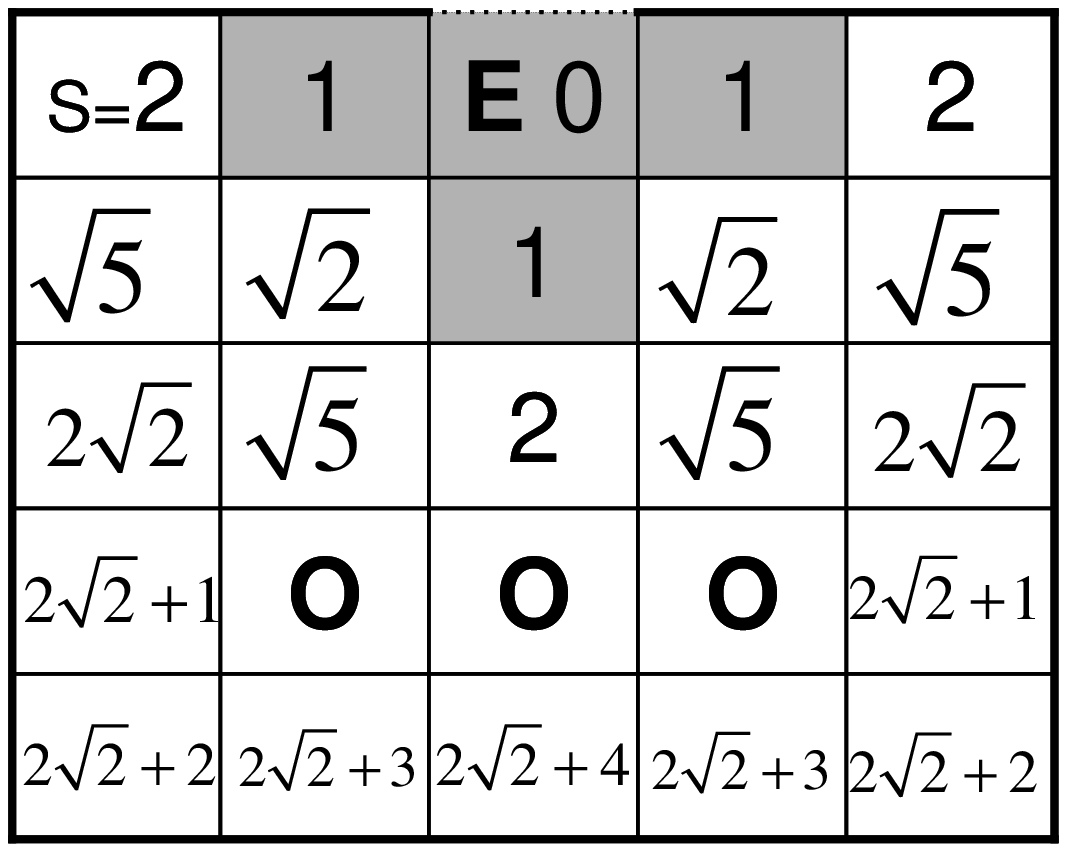}{Static floor field constructed by the exit \textbf{E}. The numbers in each cell represent the Euclidean distances from the exit cell.}{height}{3.5}}
\FIG{\EPSF{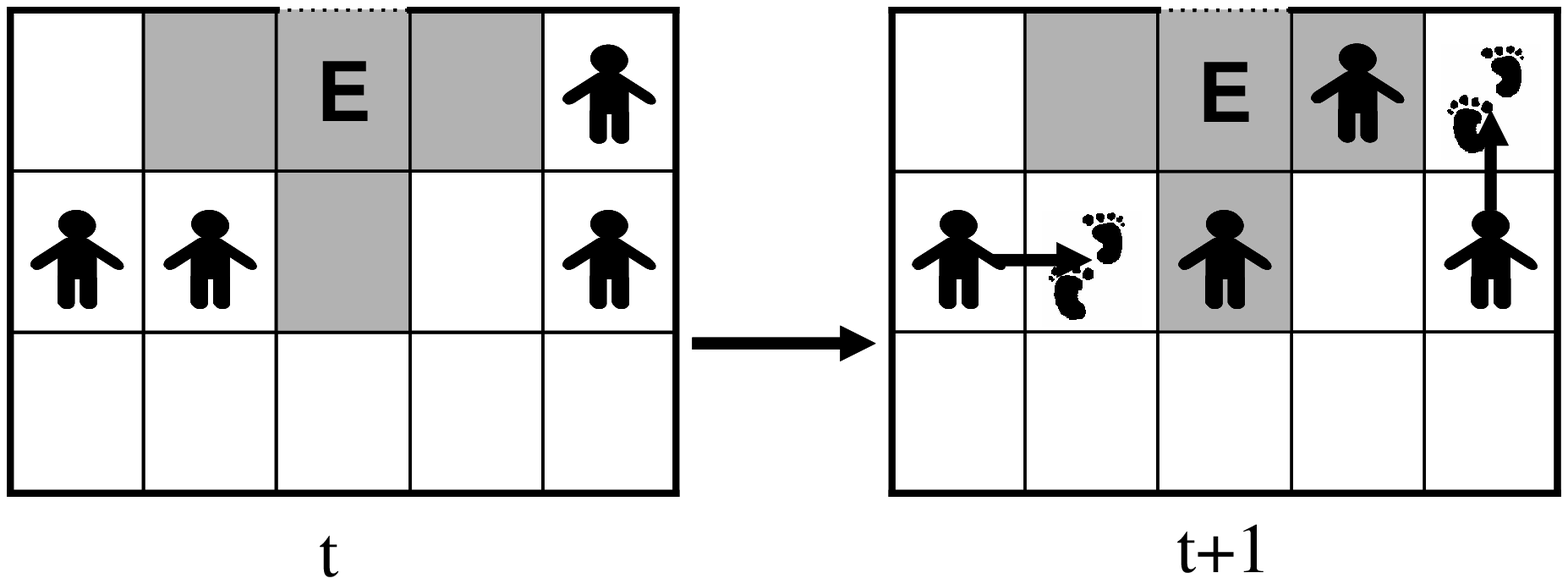}{A schematic view of dynamic floor field. At $t+1$ two pedestrians who could move leave footprints at cells where they occupied at $t$.
Remaining pedestrians are likely to move cells where footprints are left.}{height}{2.8}}

\subsection{Conflict resolution and friction} \label{collision}

Due to the use of parallel dynamics it happens that two or more pedestrians choose the same target cell in the update procedure.
Such situations are called conflicts in this paper. 
To describe the dynamics of a conflict in a quantitative way, \textit{friction parameter} $\mu\in [0,1]$ was introduced in Refs.\cite{competition,friction}.
This parameter describes clogging and sticking effects between the pedestrians.
In a conflict the movement of \textit{all} involved pedestrians is denied with probability $\mu$, i.e., all pedestrians remain at their cell.
Therefore, the conflict is solved with probability $1-\mu$, and one of the pedestrians is allowed to move to the desired cell (Fig.\ref{5}).
The pedestrian which actually moves is then chosen randomly with equal probability.
In a situation with large $\mu$ pedestrians are competitive and do not give way to others.
Thus they hardly move due to the conflict between them.
Contrary in a situation with small $\mu$ they give way and cooperate each other.

\FIGT{\EPSF{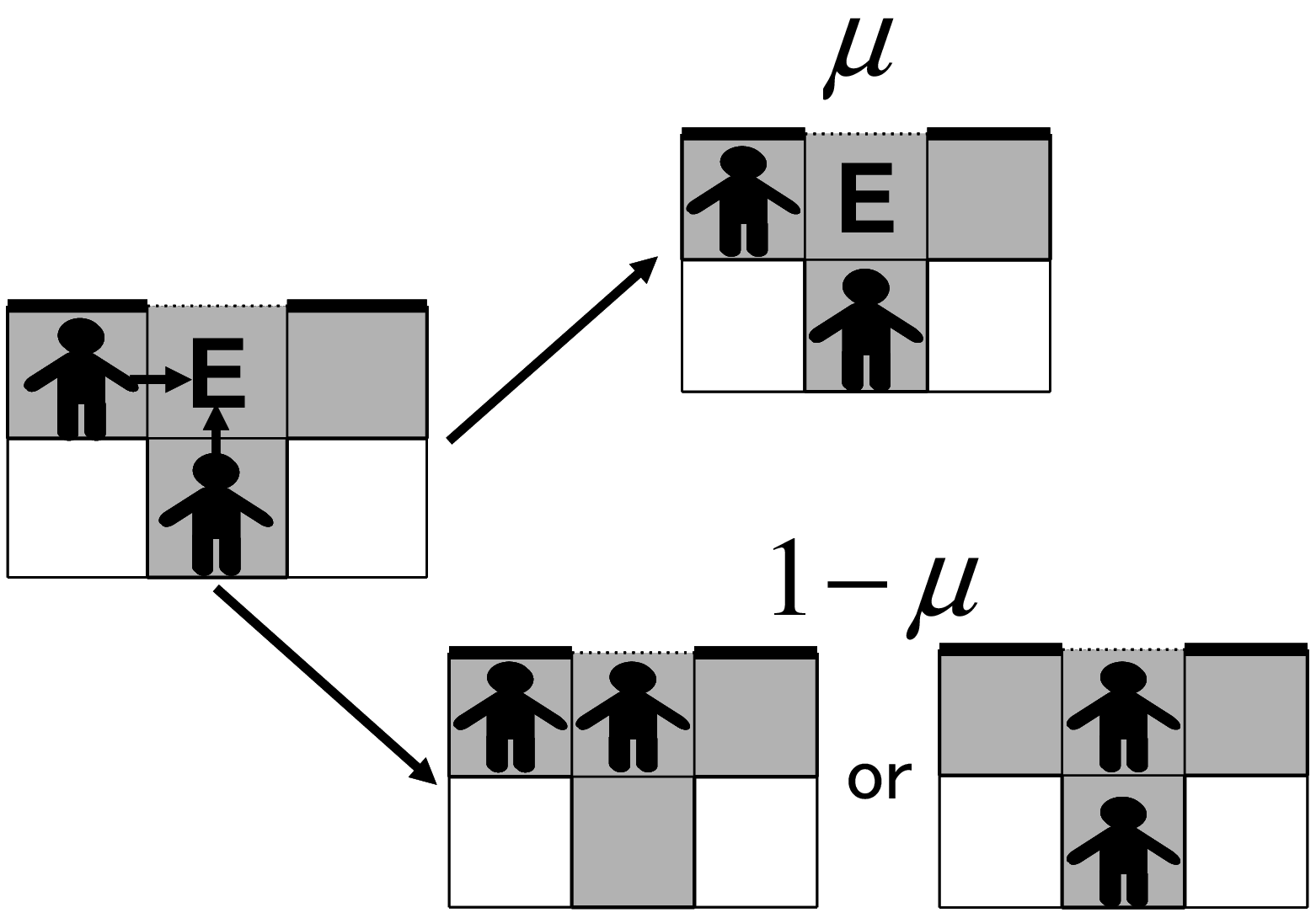}{The way of solving conflicts. In a conflict situation, movement of all involved pedestrians remain at their cell with probability $\mu$. One of them is randomly allowed to move to the desired cell with probability $1-\mu$.}{height}{4}}

\subsection{Update rules}

The FF model consists of the following 5 steps per unit time step, and is repeated until all pedestrians have exited or the maximum calculation time steps have passed.
\begin{enumerate}
\item Calculate each pedestrian's transition probability by (\ref{pp}) and the values of SFF and DFF.
\item Move pedestrians based on the calculated transition probability. If there are cells which are possibly occupied by more than two pedestrians, solve conflicts by the means of Sec.\ref{collision}.
\item Diffuse and decay DFF of every cell.
\item Pedestrians who could move at the step 2 increase a value of DFF by 1 at the cell they occupied.
\item Pedestrians who stand on exit cells are removed from the room.
\end{enumerate}
In the following, we consider only the effect of SFF and ignore DFF for simplicity.
Thus the step 3 and 4 is not considered in this paper.
DFF plays a role on mimicking the long-ranged interaction with pedestrians to short-ranged one.
Therefore, ignoring DFF is justified when we only consider pedestrian behaviors near an exit as done here.
We also confirm that results in this paper are not significantly changed by the introduction of DFF. 

\SEC{Introduction of a new parameter near the exit}{BETA}

In real situations pedestrian density depends on the area in the room.
While there are few pedestrians near the corner, there are many pedestrians gathering around the exit.
Therefore, pedestrians often conflict with each other around the exit and an arch of pedestrians is likely to be formed in front of the exit due to a friction between them \cite{socialnature}.
Figure \ref{6} shows the number of conflicts in an egress process in a competitive situation in 10,000 time steps.
An exit is set at $(x,y)=(6,10)$ in the figure. 
We see 7,842 conflicts at the exit cell, about 1,000 at the 5 Moore neighbor cells of the exit, and less than 110 at other cells.
This result says more than 60 percent of conflicts occur at the exit cell and the probability of conflicting there is about 80 percent.
Since pedestrians know this fact by experience, they walk fast when they are far from the exit, while they walk slowly or give way to each other around the exit.
That is to say, walking velocity depends on the area in the room.
In the usual FF model, however, transition probability, i.e. walking velocity is same wherever pedestrians are.
To take into account this situation, we introduced a new parameter $\beta\in[0,1]$ which we call the \textit{bottleneck parameter}.
The transition probability of pedestrians who occupy one of the Neumann neighboring cells of the exit cell is described as follows:
\FIGT{\EPSF{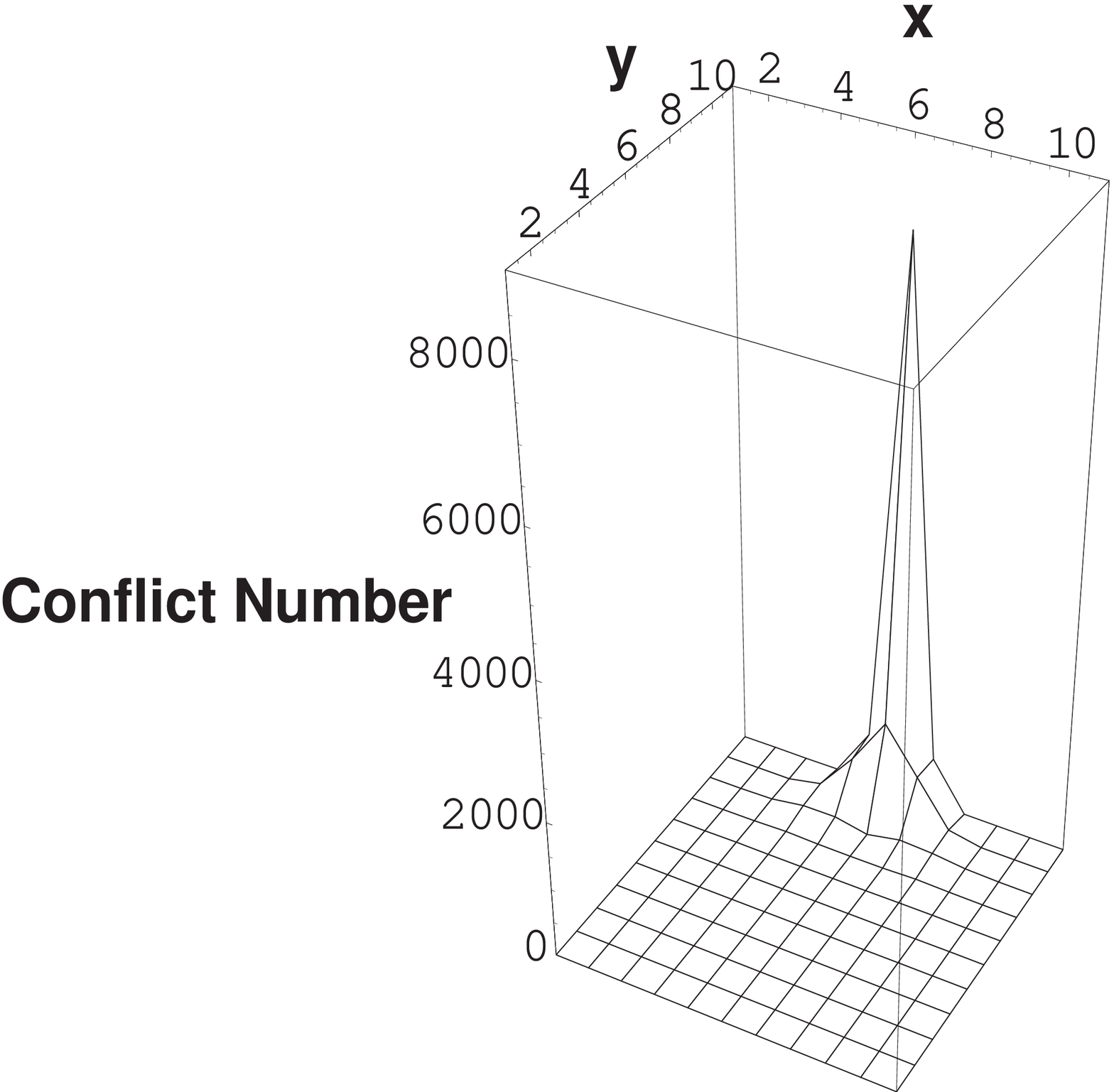}{The number of conflicts in a competitive situation. We simulated in the 11$\times$11 cells' room with the 1 cell's exit for 11,000 time steps and accumulated the value of 1,001 to 11,000 time steps. The cell at $(6,10)$ is the exit cell. We find that more than 60 percent of conflicts occur at the exit cell and the probability of conflicting there is about 80 percent.}{height}{5.2}}
\begin{eqnarray}\label{pbeta}
\left \{
\begin{array}{l}
p_{ij}=\beta \bar{N}\xi _{ij}\exp (-k_sS_{ij} +k_{d}D_{ij}) \\
\hspace{4.7cm} ((i, j)\not = (0, 0))\\
p_{0,0}=(1-\beta)+\beta \bar{N}\exp (-k_sS_{0,0} +k_{d}D_{0,0}),
\end{array}
\right.
\end{eqnarray}
where $\bar{N}$ is represented as
\begin{equation}
\bar{N} = \left[ \sum_{i,j}\xi _{ij}\exp (-k_sS_{ij} +k_{d}D_{ij})\right] ^{-1}.
\end{equation}
The transition probability of other cells is the same as (\ref{pp}).
Here, if $\beta=0$, $p_{ij}=0$ $((i, j)\not = (0, 0))$ and $p_{0,0}=1$, which means that nobody move to the exit cell.
While if $\beta=1$, the transition probability is the same as (\ref{pp}), which means that pedestrians move fast as they are far from the exit.
$\beta$ controls the velocity of the pedestrians who are at the neighboring cells of the exit.
In Ref.\cite{friction}, the parameter $k_s$ is used to describe the velocity of the pedestrians.
However, since small value of $k_s$ means lack of a knowledge of the exit position, pedestrians sometimes move backward.
In reality, pedestrians move to the exit along the shortest path and slow down near the exit to avoid conflicts with others.
Therefore, $\beta$ is not compensated by $k_s$, and we expect more realistic pedestrian behavior is seen by the parameter $\beta$.
 When $k_s$ is large, the transition probabilities of pedestrians at neighboring cells of the exit are approximated as Fig.\ref{7}. This simplification enables us to analyze the pedestrian behavior theoretically.

\FIGT{\EPSF{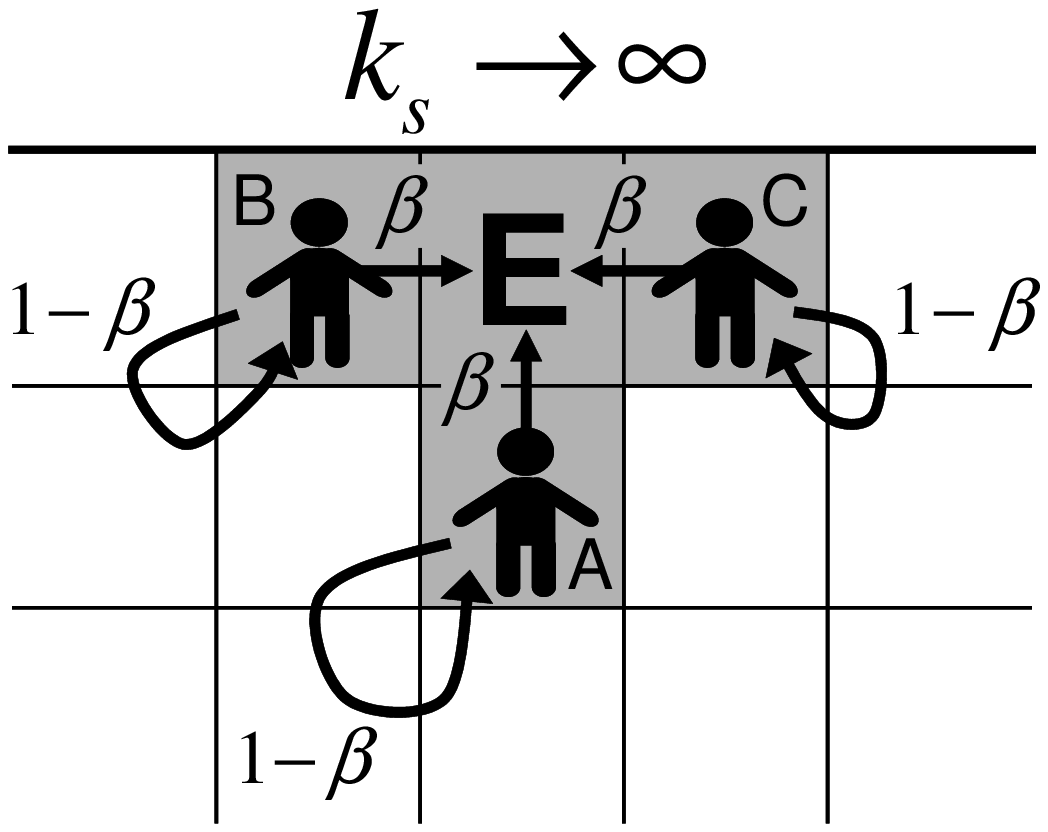}{
An approximation of the transition probabilities at Neumann neighboring cells of the exit when the width of the exit $w=1$.
When $k_s$ is large, $p_{0,1}\rightarrow \beta$, $p_{0,0}\rightarrow 1-\beta$, and $p_{0,-1}$, $p_{1,0}$,  $p_{-1,0}\rightarrow 0$ for a pedestrian A.
Similarly, $p_{1,0}\rightarrow \beta$ for a pedestrian B and $p_{-1,0}\rightarrow \beta$ for a pedestrian C.
}{height}{4.2}}

\SEC{Analytical Expression of the average flow using cluster approximation}{CLUSTER}

In the $\beta-$introduced FF model, When $k_s$ is large, it is almost sure that pedestrians move to the exit cells with the probability 1 if they are far from the exit and with the probability $\beta$ at the Neumann neighboring cells of the exit.
Therefore, in this section we focus on the exit cells and the neighbor cells of them and calculate an analytical expression of the average pedestrian flow through the exit by cluster approximation.
The flow is defined as the number of evacuated persons per 1 time step thorough an exit.
We suppose that a big jam is formed around the exit.
This enables us to simplify a situation that only SFF affect pedestrians' motion.

First, we calculate the flow when the width of the exit $w=1$.
The transition probability are defined in Fig.\ref{8}.
We consider two kinds of states of a cell 1 and 0, which represent that a pedestrian exists at the cell or not.
Therefore, in the case $w=1$, there are 16 different states for these four cells in total.
Since we assume a big jam exists at the exit, pedestrians enter into three neighboring cells of the exit with the probability 1.
$\alpha$ is the probability of getting out from the exit cell, which is set as 1 throughout this paper.
We define $p_t(0)$ as the probability that a pedestrian is not at the exit cell at time step $t$ and $p_t(1)$ as the probability that a pedestrian is at the exit cell at time step $t$. 
The master equations are described as follows:
\begin{equation}
\begin{bmatrix}
p_{t+1}(0) \\ p_{t+1}(1) \label{joutaiseni}
\end{bmatrix}
=
\begin{bmatrix}
1-r & \alpha \\ r & 1-\alpha
\end{bmatrix}
\begin{bmatrix}
p_{t}(0) \\ p_{t}(1)
\end{bmatrix}.
\end{equation}
Here $r$ represents the probability that a pedestrian enter into the exit cell from the three Neumann neighboring cells, which is described as follows:
\begin{eqnarray}
 & & {} r=\beta _1(1-\beta _2)(1-\beta _3)+\beta _2(1-\beta _3)(1-\beta _1) \nonumber\\
 & & {} \hspace{3.5cm} +\beta _3(1-\beta _1)(1-\beta _2) \nonumber\\
 & & {} \hspace{0.5cm} + (1-\mu) \{ \beta _1\beta _2(1-\beta _3)+\beta _2\beta _3(1-\beta _1) \nonumber\\
 & & {} \hspace{2.7cm}+\beta _3\beta _1(1-\beta _2)+\beta _1\beta _2\beta _3\} . \label{rnoteigi}
\end{eqnarray}
\FIGT{\EPSF{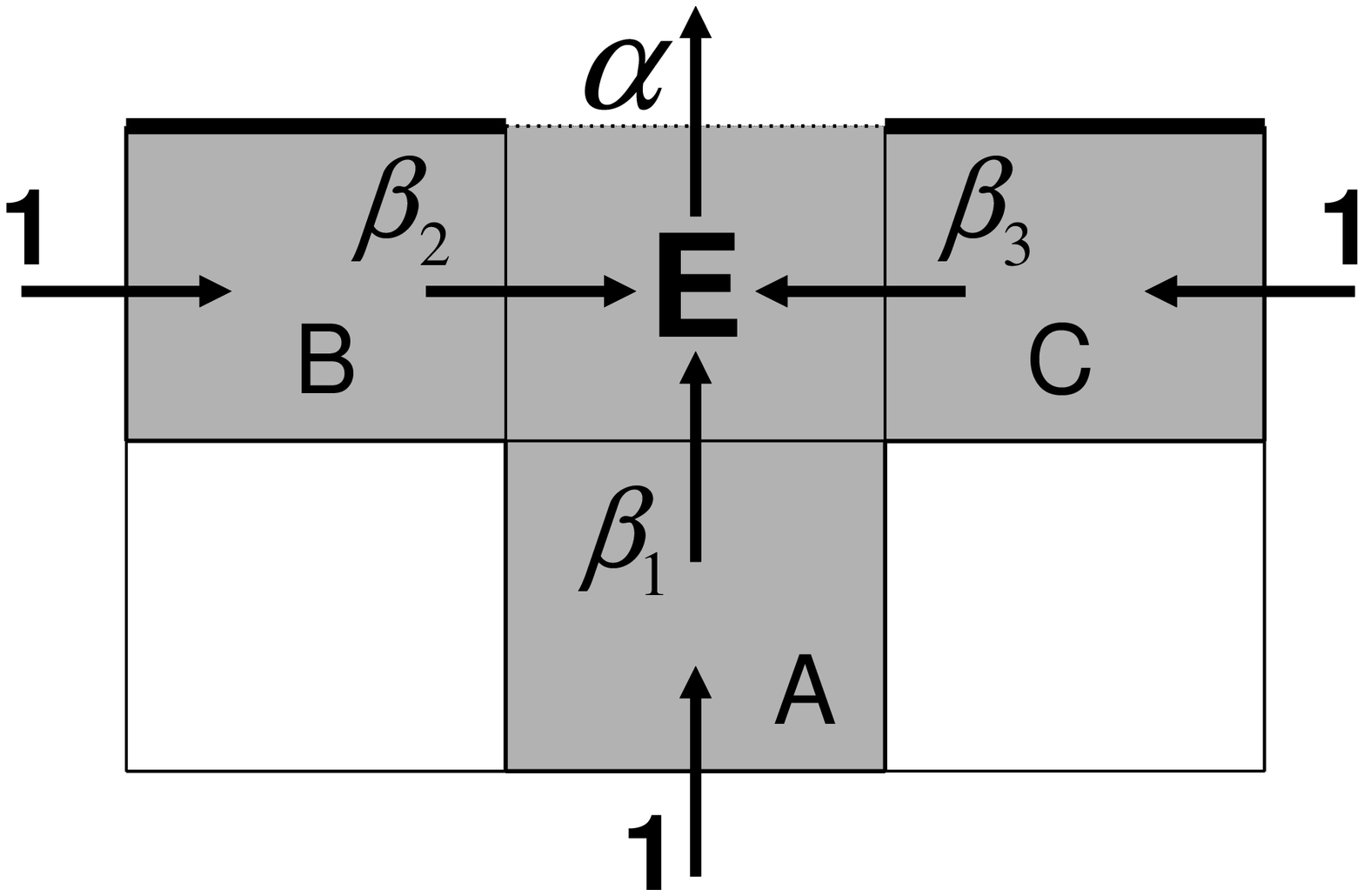}{Cluster approximation at the exit with one cell. $1$, $\alpha$, $\beta_1$, $\beta_2$, $\beta_3$ represent transition probability.}{height}{3.3}}
The first term is the probability of coming a pedestrian from the cell A (Fig.\ref{8}).
Similarly, the second and the third terms are the probability of coming a pedestrian from the cell B and C respectively.
The first three terms enclosed in parentheses at the forth term represent the probability that one of the pedestrians enter into the exit cell from two of the three cells (A, B, and C) by resolving the conflicts.
The last term enclosed in the parentheses represents a similar situation, but pedestrians enter into the exit cell from all the three neighboring cells.
By using (\ref{joutaiseni}) and (\ref{rnoteigi}) with the normalization condition
\begin{equation}
p_{t}(0)+p_{t}(1)=1,
\end{equation}
we obtain the stationary solution
\begin{equation}
p_{\infty}(1)=1-\frac{\alpha}{\alpha -a_2-a_1-a_0-\mu (a_1+2a_0)}
\end{equation}
\begin{eqnarray}
\left \{
\begin{array}{l}
a_0 = -\beta_1 \beta_2 \beta_3 \\
a_1 = \beta_1 \beta_2 + \beta_2 \beta_3 + \beta_3 \beta_1 \\
a_2 = -(\beta_1 + \beta_2 + \beta_3).
\end{array}
\right .
\end{eqnarray}
Thus the number of pedestrians who can evacuate from the exit with one cell's width per a time step, i.e. the average pedestrian flow through an exit is described as follows:
\begin{eqnarray}
\begin{split}
\langle Q(\beta_1, &\beta_2, \beta_3, \alpha, \mu) \rangle = \alpha p_{\infty}(1) \\
&= \alpha \left[ 1-\frac{\alpha}{\alpha -a_2-a_1-a_0-\mu (a_1+2a_0)} \right]. \label{Q}
\end{split}
\end{eqnarray}
The expression of the average flow in the case $\alpha=\beta_1=\beta_2=\beta_3=1$ was obtained in Ref.\cite{friction} as
\begin{equation}
\label{Nold}
\langle Q\rangle = \frac{1-\mu}{2-\mu},
\end{equation}
which can be recovered by $\langle Q(1, 1, 1, 1, \mu) \rangle$.
Thus (\ref{Q}) is a generalization of the previous result (\ref{Nold}).

Next we specify (\ref{Q}) by substituting 0 and $\beta$ for $\beta_1$, $\beta_2$, $\beta_3$ as follows:
\begin{eqnarray}
\begin{split}
(a)\ \langle q_1(\beta &, \alpha) \rangle \equiv  \langle Q(\beta, 0, 0, \alpha, \mu) \rangle  = \frac{\alpha \beta}{\alpha +\beta }, \label{q1} \\
(b)\ \langle q_2(\beta &, \alpha, \mu) \rangle \equiv  \langle Q(\beta, \beta, 0, \alpha, \mu) \rangle \\
&=\alpha \left[ 1-\frac{\alpha}{\alpha +2\beta -(1+\mu) \beta^2 } \right ], \label{q2} \\
(c)\ \langle q_3(\beta &, \alpha, \mu) \rangle \equiv  \langle Q(\beta, \beta, \beta, \alpha, \mu) \rangle \\
&= \alpha \left[ 1-\frac{\alpha}{\alpha +3\beta -3(1+\mu) \beta^2 +(1+2\mu) \beta^3 } \right ]. \label{q3}
\end{split}
\end{eqnarray}
These expressions describe average flows through an exit with the configuration described in Fig.\ref{9} respectively.
\FIGT{\EPSF{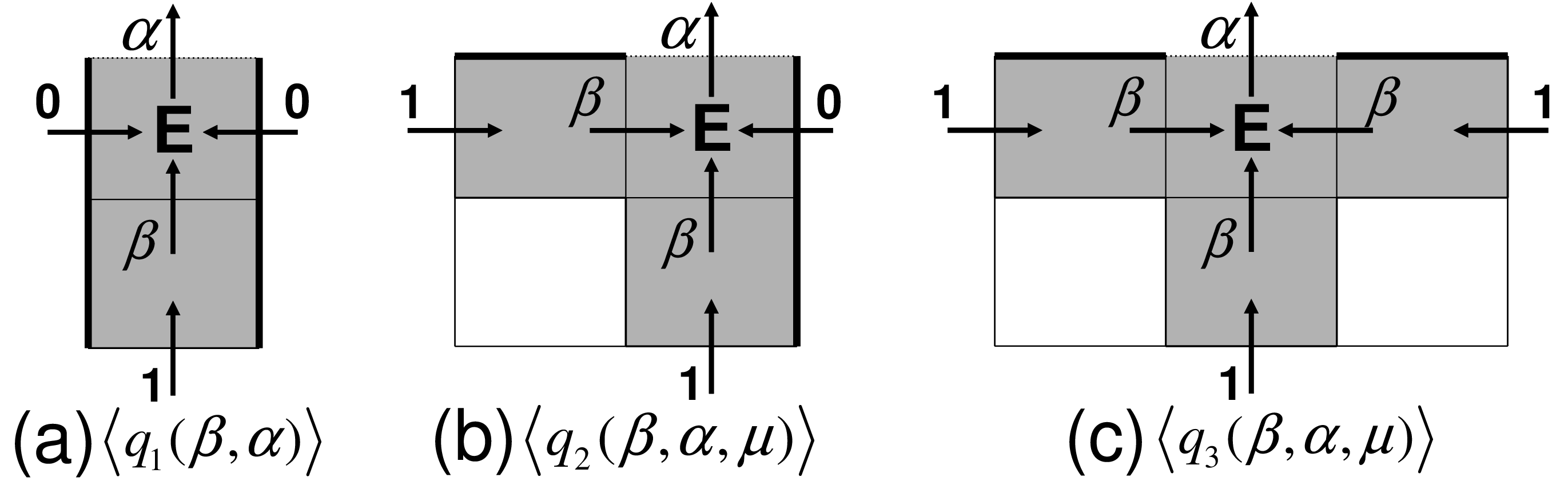}{The three special cases of Fig.\ref{8}. 
We assume that the arrow with transition probability 0 is interpreted as the existence of a wall that blocks pedestrians' motion.}{height}{2.5}}

Finally we calculate the average flow of pedestrians through an exit with arbitrary $w\in \textbf{N}$ width.
The relation between the width of an exit and the outflow is an important study which has been investigated experimentally so far \cite{ex_bottleneck1, ex_bottleneck2}.
When pedestrians move to the exit cell along the shortest path, social morals may suppress pedestrians breaking into the line.
Thus, they do not gather around the exit in disorder, but tend to form lines in front of the exit.
Moreover they do not easily change lanes in a crowd situation.
There are also the experimental results that the pedestrian outflow increases linearly as the width of an exit increases \cite{ex_bottleneck2}.
Therefore, we can represent the average flow through the exit with $w$ cell's width by linear sum of $\langle q_1 \rangle$, $\langle q_2 \rangle$, and $\langle q_3 \rangle$.
Here we consider two types of exits: an exit at the center of the wall (Ce-exit) and an exit at the corner of the room (Co-exit).
Ce-exit ($w\geq 2$) is divided into $\langle q_1 \rangle$-exits, and $\langle q_2 \rangle$-exits, and an average flow through an exit with $w$ cell's width $\langle Q_{center,w} \rangle$ is described as:
\begin{equation}
\langle Q_{center,w} \rangle =
\left \{
\begin{array}{c}
\begin{split}
\langle q_3 \rangle \ \ \ \ \ \ \ \ \ \ \ \ \ \ \ \ \ \ \ \ \ \ &(w=1) \\
2 \langle q_2 \rangle + (w-2) \langle q_1 \rangle \ \ &(w \geq 2).
\end{split}
\end{array}
\right.
\label{center}
\end{equation}
In a similar way, Co-exit is divided into $\langle q_1 \rangle$-exits, and $\langle q_2 \rangle$-exit, and an average flow through an exit with $w$ cell's width $\langle Q_{corner,w} \rangle$ is described as follows:
\begin{equation}
\langle Q_{corner,w} \rangle = \langle q_2 \rangle + (w-1) \langle q_1 \rangle \ \ \ (w \geq 1).
\label{corner}
\end{equation}
The examples of dividing Ce-exit and Co-exit ($w=3$) are shown in Fig.\ref{10}.
We also define the average flow per 1 cell as:
\begin{equation}
\left \{
\begin{array}{c}
\langle q_{center} \rangle = \langle Q_{center,w} \rangle / w \\
\langle q_{corner} \rangle = \langle Q_{corner,w} \rangle / w.
\end{array}
\right.
\end{equation}

\FIGT{\EPSF{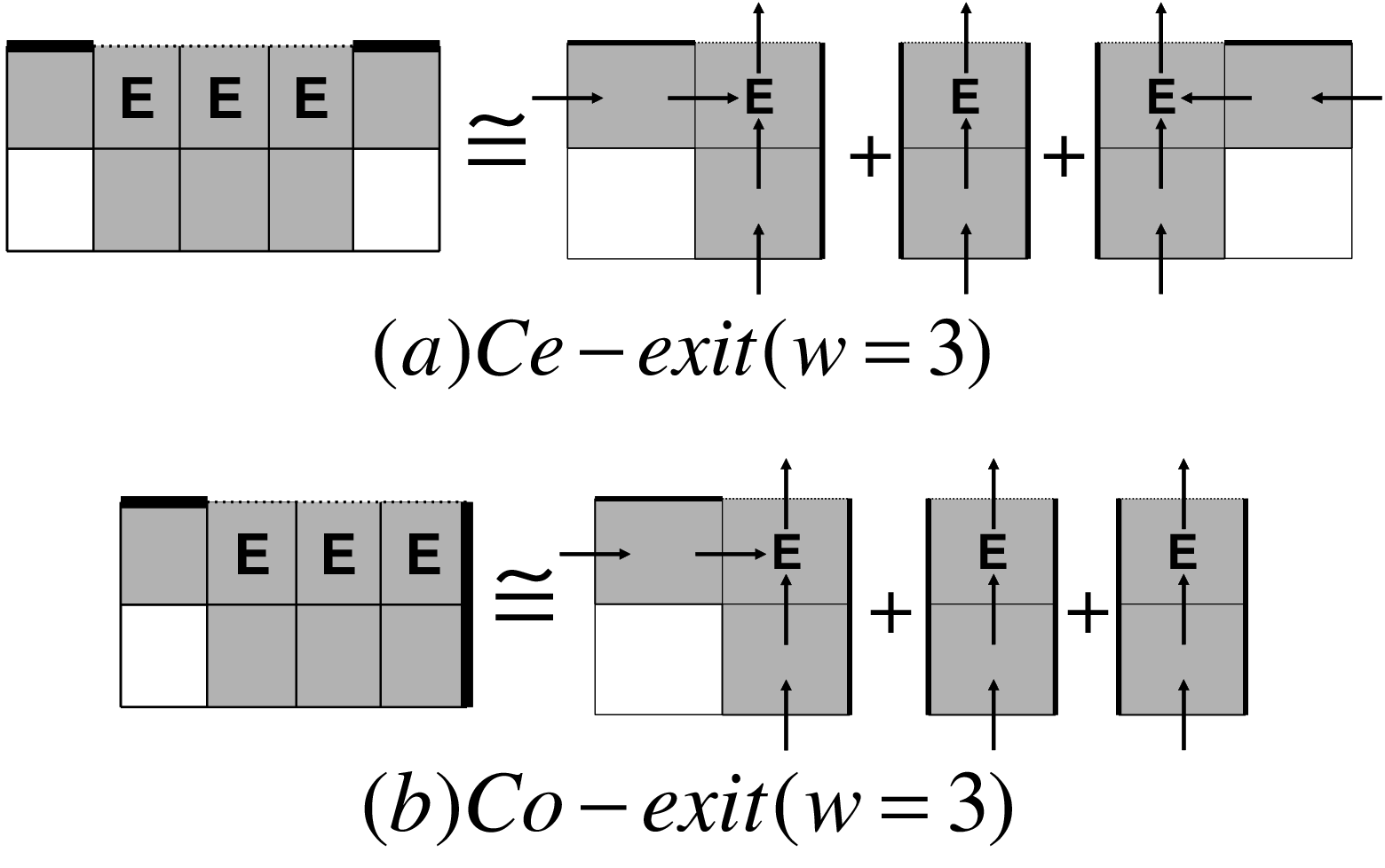}{Dividing an exit with 3 cells into three exits with 1 cell. The outflow is also represents the sum of each flow.}{height}{4.5}}

\SEC{Comparing the average flows of analysis and simulation}{S&E}

In this section we compare the analytical and computational results of the average flow $\langle q \rangle$, which is a function of $\beta$.
The parameters of FFs are set at: $k_s=10$ and $k_d=0.0$ so that pedestrians move to the exit along the shortest path. 
The size of the room used in simulation is 11$\times$11 cells.
Ce-exit room has the exit cells at the center of the boundary cells of one side of the room, and those of the other sides are all entrance cells where pedestrians come with the probability 1.
Similarly, Co-exit room has the exit cells at the corner of the room, and boundary cells of two sides which do not include the exit cells are all entrance cells.
The examples of the 5$\times$5 rooms are shown in Fig.\ref{11}.
We simulated 11,000 time steps with the initial condition that pedestrians occupy all cells except exit and obstacle cells.
Then the average flow of 10,000 time steps from 1,001 to 11,000 is used to depict Fig.\ref{12}.
It shows average pedestrian flows at the exit as a function of $\beta$ for various $\mu$ values.
We see that the simulations agree with the analytical results very well. 
The errors become large for $\mu = 0.9$ since pedestrians conflict with each other and then they cannot move not only at the exit cells but also at the other cells.
Surprisingly, for $\mu = 0.9$, we clearly find that a maximum flow is attained at a value of $\beta$ in both simulations and analytical results.
We call it as the optimal $\beta$, notated as $\beta_{opt}$ here after.
For $\mu \rightarrow 1$ the number of unsolved conflicts increases as $\beta$ grows.
As a result, pedestrians stick and the average pedestrian flow decreases.
We also find that the differences of the flows for different $\mu$ is getting smaller as $w$ increases by comparing the lengths of the arrows in Fig.\ref{12}, i.e., the arrow C is shorter than A and D is shorter than B.
For the same $w$ the differences of the flows for different $\mu$ are smaller at the Co-exit than the Ce-exit since the arrow B is shorter than A and D is shorter than C.

\FIGT{\EPSF{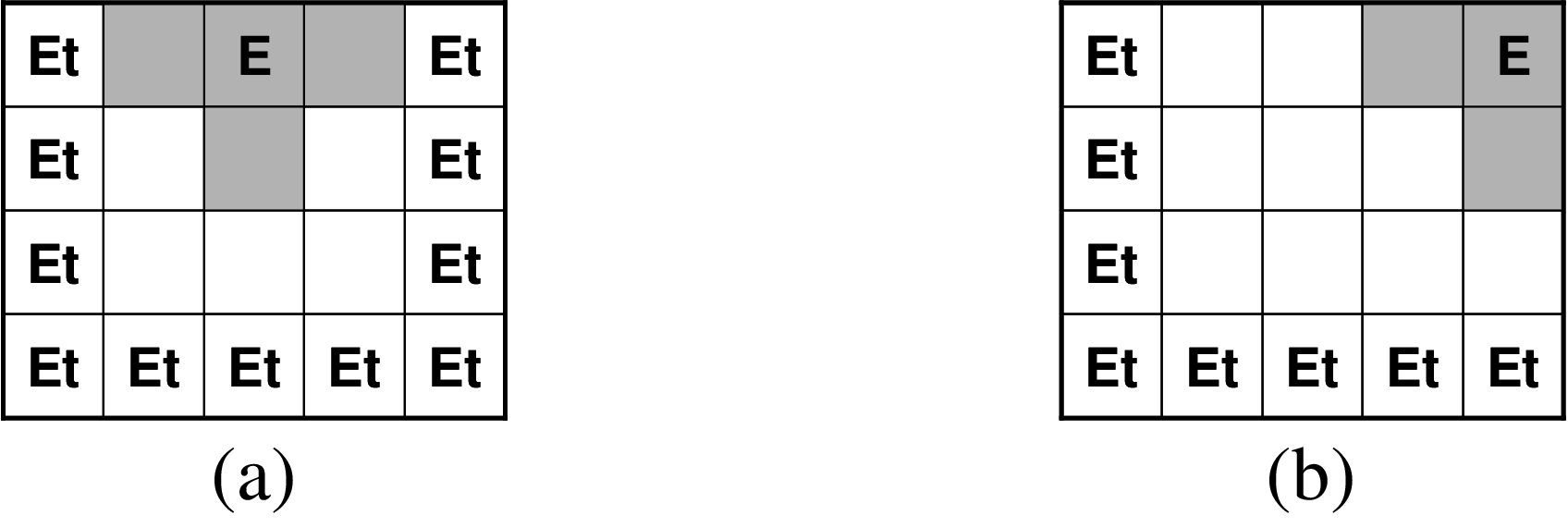}{$5\times 5$ cells' rooms with 1 cell's exit. (a)Ce-exit. (b)Co-exit. \textbf{E} represents an exit cell, and \textbf{Et} represents an entrance cell where pedestrians come with probability 1.}{height}{2.5}}

\FIGT{\EPSF{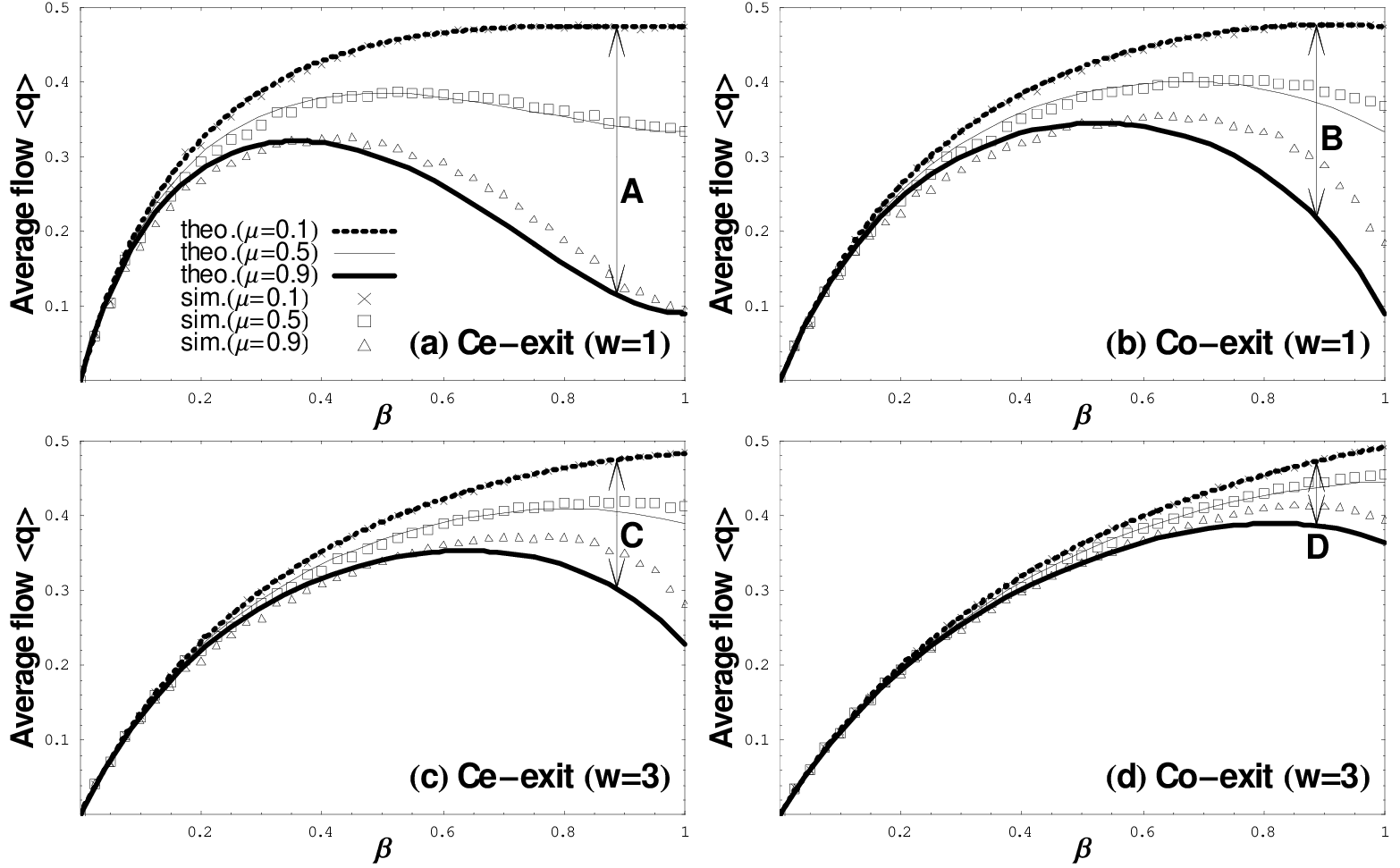}{
Average flow $\langle q \rangle$ as a function of $\beta$ for different $\mu$, exit width $w$, and exit location.
(a)Ce-exit, $w=1$. (b)Co-exit, $w=1$. (c)Ce-exit, $w=3$. (d)Co-exit, $w=3$.
For $\mu=0.9$ we clearly see a maximum of flow at an optimal $\beta$.
The lengths of the arrows A, B, C and D represent the differences of the flows for different $\mu$.}{height}{5.5}}

\SEC{Competitive, cooperative behavior and the width of an exit}{C&C}

Kirchner \textit{et al} made a research on how the pedestrians' mood influences on the evacuation time \cite{competition}.
Both the experimental and the computational results show that competition is beneficial if the exit width exceeds a certain width, and harmful if the exit width is lower than it.
We explain this phenomenon by the analytical solutions (\ref{center}) and (\ref{corner}).
In Ref.\cite{competition}, competition is described as a increased assertiveness (large $k_s$) and a strong hindrance in conflict situations (large $\mu$).
Cooperation is represented by small $k_s$ and vanishing $\mu$.
In our new Model we describe the assertiveness by $\beta$.
We use $\mu$ as a parameter of hindrance in conflict situations and its values are same as in Ref.\cite{competition}.
The parameters are set at $\beta=1.0$, $\mu=0.6$ for the competitive situation and $\beta=0.4$, $\mu=0.0$ for the cooperative situation.

Figure \ref{13} shows the average flows for variable door width $w$.
The results of analysis agree with those of simulation very well.
The simulation condition is the same as we described in Sec.\ref{S&E}; however, we used 12$\times$12 rooms to set up the exit at the center of the boundary cells of the room if the width of the exit is even number.
The size of the room does not influence the average flow since most of the cells are occupied by pedestrians.
Clearly we can observe the crossing of the two curves at a critical door width $w_c\approx 3$ in Fig.\ref{13}(a).
Our result is well correspond to the Ref.\cite{competition}'s result, which is $w_c\approx 2.5$.
This means that we should cooperate with each other to increase the average pedestrian flow when the width of the exit is narrow.
On the contrary, when the width of the exit is wide, we do not have to give way to other pedestrians and should go through the exit aggressively.
When the exit door is at the corner of the room, the crossing is observed at $w_c\approx 1.5$ in Fig.\ref{13}(b).
Therefore, Co-exit is more suitable for competitive situation than Ce-exit.

The Japanese building standards law \cite{kenchiku} gives an average pedestrian flow 1.5[persons/(m$\cdot$s)] if an exit is directly connected to the ground.
We find that this value significantly changes by the pedestrians' moods, i.e., competitive and cooperative.
From Fig.\ref{13} we obtain the values of the average flow through Ce-exit i.e., 1.5[persons/(m$\cdot$s)] in the  competitive situation and 2.0[persons/(m$\cdot$s)] in the cooperative situation.
The values are calculated by defining the cell size as 50[cm]$\times$50[cm] and using a pedestrian velocity 1.3[m/s] which is according to the Japanese building standards law.

\FIGT{\EPSF{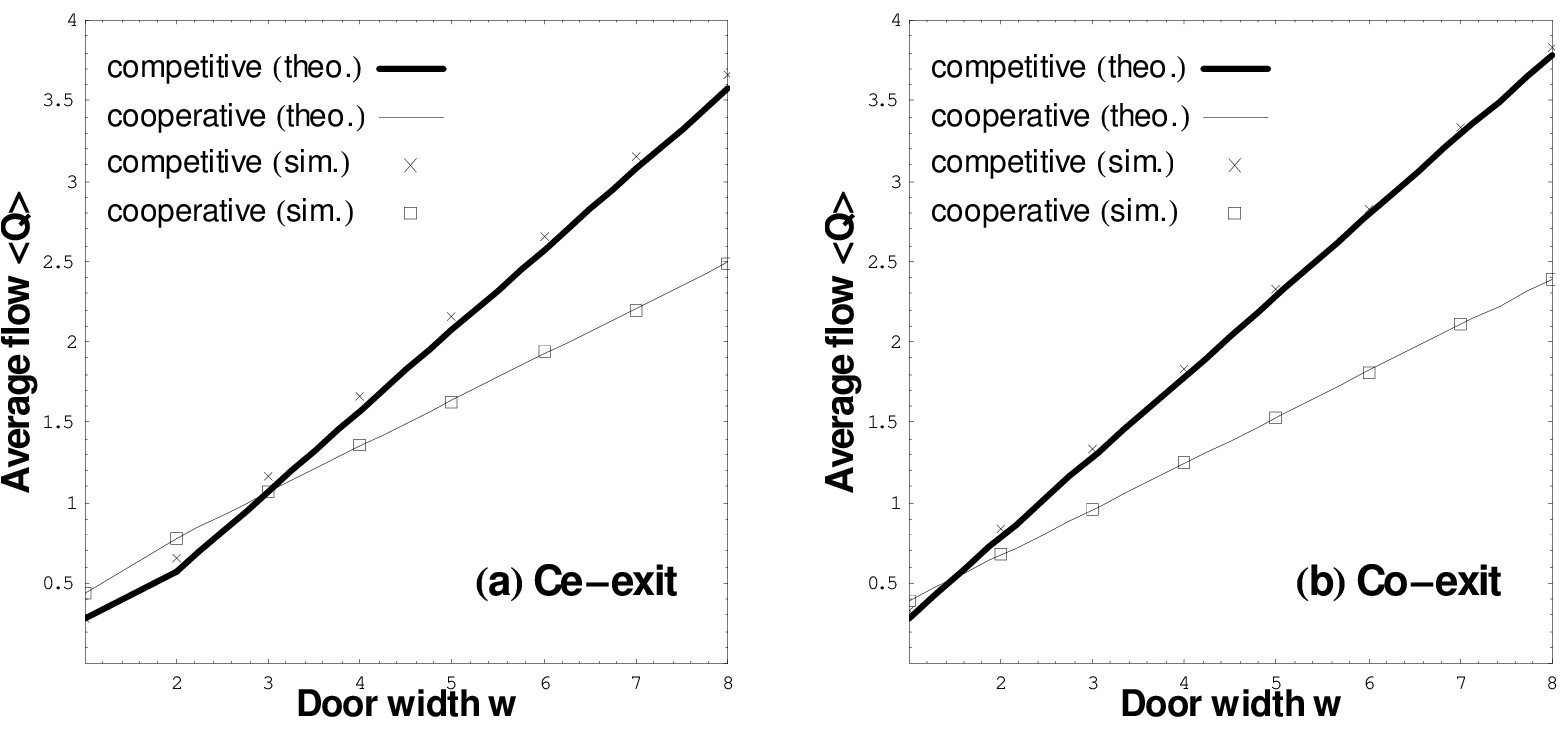}{Average flow for various exit door width $w$. (a)Ce-exit. (b)Co-exit.
We observe the crossing of the two curves in both (a) and (b).}{height}{4.2}}

\SEC{Competitive, cooperative behavior and the effect of a wall}{WALL}

Here we compare the average flows of Ce-exit and Co-exit, and discuss how the wall has an effect on them.
The difference of $\langle Q_{center,w} \rangle$ and $\langle Q_{corner,w} \rangle$ is calculated as follows:
\begin{eqnarray}
\label{Qdifference}
\begin{split}
\langle Q_{corner,w} \rangle &- \langle Q_{center,w} \rangle \\
&=
\begin{cases}
-\beta (\beta -\frac{1}{1+2\mu})(\beta -1)A & (w=1) \\
\beta (\beta-\frac{1}{1+\mu})B & (w \geq 2),
\end{cases}
\end{split}
\end{eqnarray}
where $A$ and $B$ are positive in the entire domain of $\beta$ and $\mu$ that is described as
\begin{eqnarray}
\begin{split}
A&=\frac{\alpha ^2(1+2\mu )}{\{ \alpha +2\beta -(1+\mu ) \beta^2 \} } \\
 & \hspace{1.0cm} \cdot \frac{1}{\{ \alpha +3\beta -3(1+\mu) \beta^2 +(1+2\mu) \beta^3 \} } \\
B&=\frac{\alpha ^2(1+\mu)}{(\alpha + \beta )\{ \alpha +2\beta -(1+\mu ) \beta^2 \} }.
\end{split}
\end{eqnarray}
We obtain $\beta_{c}$ which is the value of $\beta$ that $\langle Q_{center,w} \rangle$ equals $\langle Q_{corner,w} \rangle$ as follows:
\begin{equation}
\label{betawall}
\beta_{c} =
\begin{cases}
\frac{1}{1+2\mu} & (w=1) \\
\frac{1}{1+\mu} & (w \geq 2),
\end{cases}
\end{equation}
The curves of (\ref{betawall}) are drawn in Fig.\ref{14}.
They divide the $\beta-\mu$ plane into two regions.
In the lower left region, the Ce-exit flow is larger and in the upper right region the Co-exit flow is larger.
We also plot the competitive and cooperative situation used in Sec.\ref{C&C}.
The figures show that the Co-exit flow is larger in the competitive situation since the wall prevents pedestrians rushing to the exit at the same time, but the Ce-exit flow is larger in the cooperative situation.
From this result, we can say that an exit should be at the center of a wall when pedestrians are in the cooperative mood, and should be at the corner of the room when people are in the competitive mood.

\FIGT{\EPSF{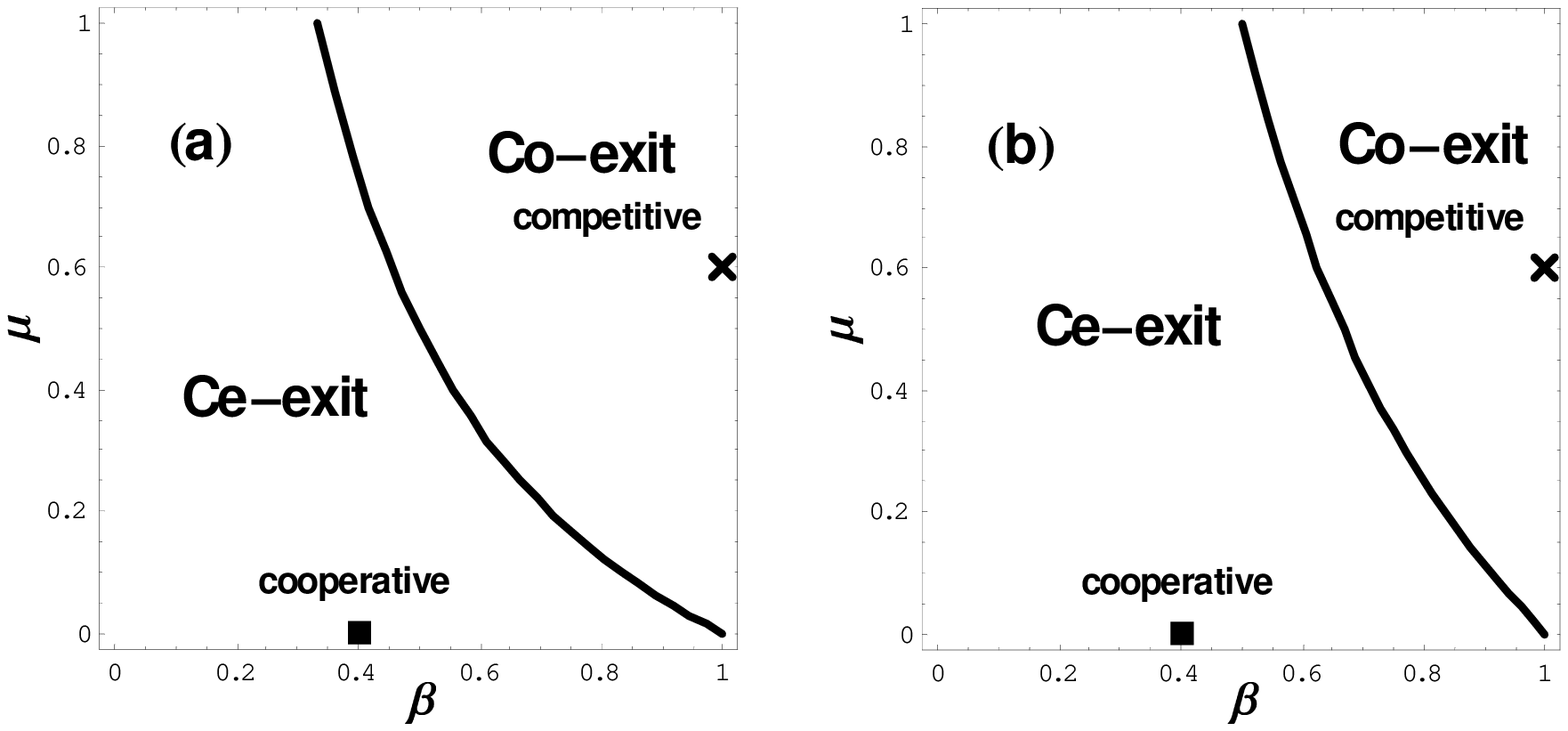}{The curves of $\beta_c$ on $\beta-\mu$ plane. (a)$w=1$. (b)$w \geq 2$. The flow of Co-exit is larger than the flow of Ce-exit in the upper right region in the figures. 
In the lower left region, the Ce-exit flow is larger.}{height}{4.1}}

\SEC{A change of contour plots of the average flow}{CONTOUR}

The average pedestrian flow through an exit is decided by three parameters: $\alpha$, $\beta$, and $\mu$ according to (\ref{Q}).
Since we have set at $\alpha=1$ in this paper, the average flow is determined by $\beta$ and $\mu$.
Fig.\ref{zubeta-mu} are contour plots of the average flow, in terms of $\beta$ and $\mu$.
Fig.\ref{zubeta-mu} (a), (b), (c), and (d) correspond to Fig.\ref{12} (a), (b), (c), and (d) respectively.
The flow is large in the white region and small in the black region.
The values of the flow are normalized in each figure for drawing the gray scaled figures.
The thick curves in the figures are the curves of $\beta_{opt}$, which gives a maximum average flow in the case of a constant $\mu$.
$\beta_{opt}$ curves divide the plane into two regions.
In the upper right regions pedestrians should slow down further to avoid a conflict and in the lower left regions they have to speed up to increase the average flow around the exit.
We calculate $\beta_{opt1}$, $\beta_{opt2}$, and $\beta_{opt3}$ from $\langle q_1 \rangle$, $\langle q_2 \rangle$, and $\langle q_3 \rangle$ given in (\ref{q2}), respectively as follows:
\FIGT{
{\ING{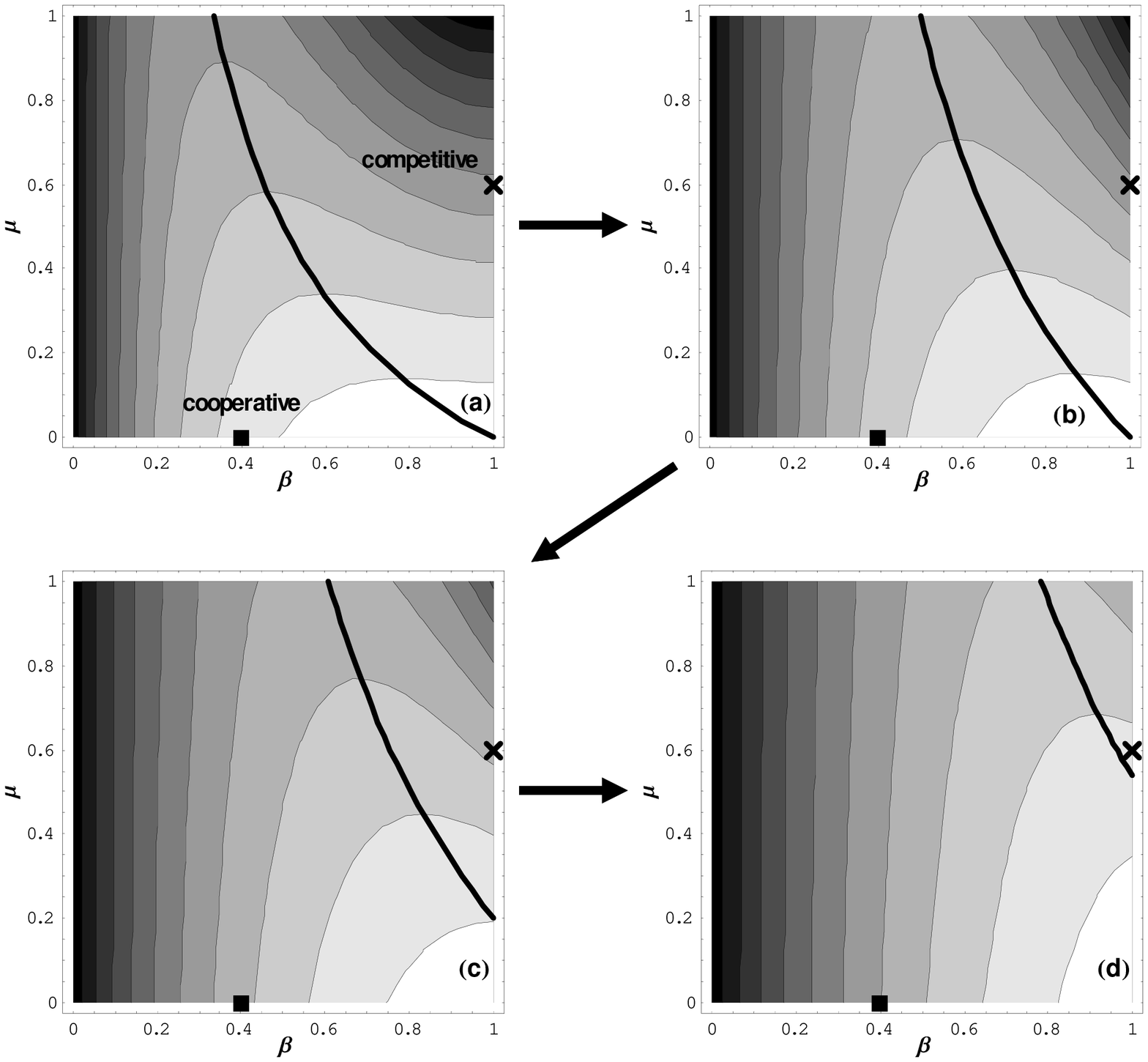}{height}{8.2}}
\NAME{zubeta-mu}{Contours of average flows for various width of the exit and its position. (a)Ce-exit, $w=1$. (b)Co-exit, $w=1$. (c)Ce-exit, $w=3$. (d)Co-exit, $w=3$.
\textbf{×} represents competitive situation and ■ represents cooperative situation given in Sec.\ref{C&C}.
Black bold curves represent $\beta_{opt}$, which move to the right in order (a), (b), (c), and (d).  
In Fig.(a) and (b) flow in the cooperative situation is larger than that of the competitive situation. 
In Fig.(c) the average flows are almost the same in both the competitive and the cooperative situation.
In Fig.(d) flow in the competitive situation is larger.
These figures explain that flow in the competitive situation is getting larger than that in the cooperative situation by increasing the width of the exit and the effect of wall.}
}
\begin{equation}
\begin{split}
\label{betaopt}
\beta_{opt1} &= 1, \\
\beta_{opt2} &= \frac{1}{1+\mu}, \\
\beta_{opt3} &= \frac{1}{1+2\mu}.
\end{split}
\end{equation}
The $\beta_{opt}$ of $\langle Q_{center,w} \rangle$ and $\langle Q_{corner,w} \rangle$ are straight-forward, but not expressed in a simple form, so we omit them in this paper.
We also plot the competitive and cooperative situations used in Sec.\ref{C&C} in the figures.

We see that the $\beta_{opt}$ curves move to the right in the order: (a), (b), (c), and (d). 
This is explained by (\ref{center})，(\ref{corner})，and (\ref{betaopt}).
First, we compare the $\beta_{opt}$ curve's position of $\beta_{opt1}$, $\beta_{opt2}$, and $\beta_{opt3}$ in the $\beta-\mu$ plane by (\ref{betaopt}).
The curve of $\beta_{opt3}$ is at the most left, that of $\beta_{opt2}$ is in the middle, and that of $\beta_{opt1}$ is at the most right (Fig.\ref{16}).
Next, the expressions of the average flow corresponding to (a), (b), (c), and (d) are described as follows:
\begin{eqnarray} \label{Qabcd}
\begin{split}
Q(a)&\equiv \langle Q_{center}(w=1) \rangle = \langle q_3 \rangle \\
Q(b)&\equiv \langle Q_{corner}(w=1) \rangle = \langle q_2 \rangle \\
Q(c)&\equiv \langle Q_{center}(w=3) \rangle = 2\langle q_2 \rangle + \langle q_1 \rangle \\
Q(d)&\equiv \langle Q_{corner}(w=3) \rangle = \langle q_2 \rangle + 2\langle q_1 \rangle
\end{split}
\end{eqnarray}
Now we see clearly why the $\beta_{opt}$ curve of $Q(b)$ is at more right than that of $Q(a)$.
$Q(c)$ includes $\langle q_2 \rangle + \langle q_1 \rangle$ more than $Q(b)$, and $Q(d)$ includes $2\langle q_1 \rangle$ more than $Q(b)$.
Therefore, $\beta_{opt}$ curve moves to the right in order (a), (b), (c), and (d).

In the lower left regions of the figures the average flow increases as $\beta$ increases, but in the upper right regions it decreases as $\beta$ increases for fixed $\mu$.
Thus the flow increasing region expands as $\beta_{opt}$ curves move to the right.
This makes an exit more suitable to the competitive situation than the cooperative situation.
We can interpret that increasing of the width of a exit and the effect of a wall make the average flow larger in competitive situation than cooperative situation, since $\beta_{opt}$ curves move to the right by both increasing of the width of a exit ((a)→(c)，(b)→(d)) and the effect of a wall ((a)→(b)，(c)→(d)).

\FIGT{\EPSF{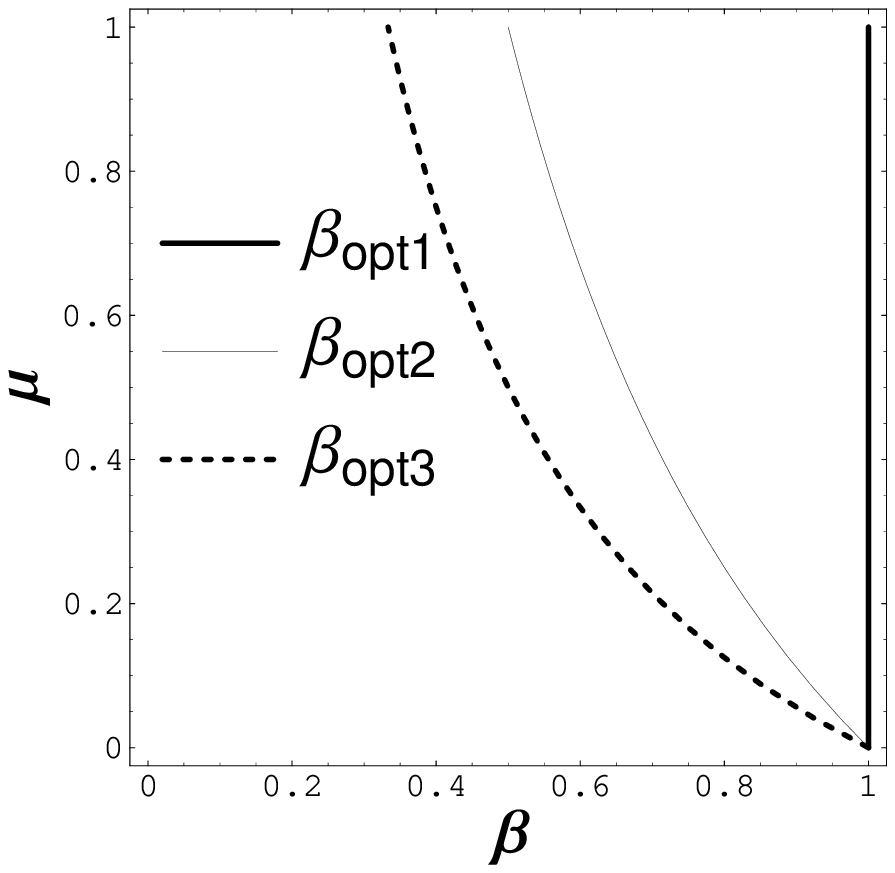}{The curve of $\beta_{opt1}$, $\beta_{opt2}$, and $\beta_{opt3}$.
We see that the curve of $\beta_{opt3}$ is at the most left, that of $\beta_{opt2}$ is in the middle, and that of $\beta_{opt1}$ is at the most right.}{height}{4.5}}

\SEC{Conclusion}{CONC}

We have introduced it to the FF model that the effect of slowing down of pedestrians around an exit, and obtained the analytical expression of the average flow through an exit with arbitrary $w$ cells by employing cluster approximation.
It turns out that the theoretical results agree quite well with the simulations.
The effects of pedestrians' mood, a width of an exit, and wall effect are also studied.
The critical exit door width, which was first obtained experimentally and was reproduced by simulations in Ref.\cite{competition}, is also analytically obtained in this paper.
We find that an exit should be at the center of a wall in the cooperative situation whereas it should be at the corner of the room in the competitive situation for smooth evacuation. 
The theoretical results also tell us that the unsolved conflicts between pedestrians around the exit are the main cause of decrease in the average pedestrian flow.
Therefore, we should consider how to decrease conflicts at a bottleneck to get large pedestrian outflow.

It is important to study pedestrian's behavior quantitatively by theoretical analysis, since its dynamics are mainly studied by simulations so far. 
The Japanese building standards law gives the average pedestrian flow through an exit as a constant value 1.5[persons/m$\cdot$s] \cite{kenchiku}.
Our expression of the average pedestrian flow is more precise and realistic, thus our results can be applied to the design of buildings so that pedestrians evacuate safely and quickly.
For example, many present concert halls have an exit at the center of the wall, however, according to our study we can shorten an evacuation time by setting up an exit at the corner of the hall when people rush into the exit in competitive mood.

In this paper we consider the average flow through the exit with more than 1 cell as the linear sum of the flow through an exit with 1 cell. 
In the calm situation, social morals suppress pedestrians cutting into lines, however, in the panic situation, they break into lines to save their lives.
Introducing such interactions between neighboring cells of an exit in detail is the future works.

\section*{Acknowledgments}
We thank Andreas Schadschneider, Armin Seyfried and Christian Rogsch for the useful discussions.

\end{document}